\documentclass{comjnl}

\usepackage{amsmath}

\usepackage{graphicx,verbatimbox}

\usepackage{makecell}
\usepackage[utf8]{inputenc}
\usepackage{caption}

\usepackage[hyphens]{url}

\usepackage[draft] {hyperref}

\usepackage{cite}
\usepackage [english]{babel}
\usepackage [autostyle, english = american]{csquotes}
\MakeOuterQuote{"}

\begin{document}

\title[CertLedger: A New PKI Model with Certificate Transparency Based on Blockchain]{CertLedger: A New PKI Model with Certificate Transparency Based on Blockchain}

\author{Murat Yasin Kubilay}
\affiliation{Department of Computer Engineering, Gebze Technical University, Kocaeli, TR} 
\author{Mehmet Sabır Kiraz}
\affiliation{TÜBİTAK BİLGEM, Kocaeli, TR} 
\author{Hacı Ali Mantar}
\affiliation{Department of Computer Engineering, Gebze Technical University, Kocaeli, TR} \email{muratkubilay@gtu.edu.tr\\mehmet.kiraz@tubitak.gov.tr\\hamantar@gtu.edu.tr}

\shortauthors{M. Y. Kubilay, M. S. Kiraz, and H. A. Mantar}



\begin{abstract}

In conventional PKI, CAs are assumed to be fully trusted. However, in practice, CAs' absolute responsibility for providing trustworthiness caused major security and privacy issues. To prevent such issues, Google introduced the concept of \emph{Certificate Transparency (CT)} in 2013.  Later, several new PKI models (e.g., AKI, ARPKI, and DTKI) are proposed to reduce the level of trust to the CAs. However, all of these proposals are still vulnerable to split-world attacks if the adversary is capable of showing different views of the log to the targeted victims. In this paper, we propose a new PKI architecture with certificate transparency based on blockchain, what we called \emph{CertLedger}, to eliminate the split-world attacks and to provide an ideal certificate/revocation transparency. All TLS certificates, their revocation status, entire revocation process, and trusted CA management are conducted in the CertLedger. CertLedger provides a unique, efficient, and trustworthy certificate validation process eliminating the conventional inadequate and incompatible certificate validation processes implemented by different software vendors. TLS clients in the CertLedger also do not require to make certificate validation and store the trusted CA certificates anymore. We analyze the security and performance of the CertLedger and provide a comparison with the previous proposals.

\end{abstract}

\maketitle

\section{Introduction}

\label{intro}
Web based applications such as internet banking, mailing, e-commerce are playing a major role for facilitating our life and became an indispensable part of it.  As a de facto standard, SSL/TLS certificates are used to provide authenticity, integrity and confidentiality services to these applications. These certificates are issued by CAs which are assumed to be trusted organizations in the conventional PKI systems. In particular, CAs are expected to operate according to some rules which are announced as Certificate Policy (CP) and Certificate Practice Statement (CPS) documents. In the current trust model, CAs have the absolute responsibility to issue correct certificates for the designated subject.  However, CAs can still be compromised and fake but valid certificates can be issued due to inadequate security practices or non-compliance with the CP and CPS. During the last decade, there have been serious incidents due to aforementioned reasons which we shortly mention below.

\begin{itemize}
	\item  The Stuxnet \cite{falliere2011w32} malware is signed by the private keys of two compromised Taiwanese CAs which targets to control a specific industrial system likely in Iran, such as a gas pipeline or power plant.
	
	\item Comodo CA, which has a big share in SSL market is hacked in March 2011\cite{Comodo2011}. One of its Registration Authority (RA) is attacked to issue 9 fraudelent certificates where the attacker is traced back to Iran. 
	
	\item A Dutch CA DigiNotar is pawned in July 2011\cite{DigiNotar2011}. 531 fraudulent certificates are issued for valuable domains such as *.google.com, *.windowsupdate.com and *.mozilla.com. These certificates could easily be used to distribute malicious Windows updates or Firefox plugins without taking attention. At least 300.000 unique IPs are detected using Google services through these certificates, which 99\%  of the traffic is from Iran. 
	
	\item Trustwave CA has sold a subCA certificate for one of its subordinates. This subCA has issued  fraudelent TLS certificates which are used to introspect TLS traffic\cite{Trustwave2012}. 
	
	\item A Turkish CA Turktrust has mistakenly issued  subCA certificates  instead of TLS certificates in December 2012\cite{Turktrust2012}. These certificates are used to generate TLS certificates for traffic introspection. Google identified the fraudulent Google certificate via Chrome.
	
	\item A subCA of the Chineese CA CNNIC, which is located in Egypt, issued fraudulent TLS certificates for traffic introspection in March 2015\cite{CNNIC2015}. Later on, it is identified that CNNIC is operated without documented CPS. 
	
	\item Lenova Superfish has deployed local CA in its products in 2015\cite{Lenovo2015}. This CA is used to inject ads into the TLS protected web sites. Since the CA private keys are in the computer RAM, they may be easily used to introspect traffic.
	
	\item Symantec issued unauthorized certificates for Google domains in September 2015 \cite{Symantec2015}. Later on Symantec claimed that these certificates are produced for test purposes.
	
	\item Symantec purchased Blue Coat in May 2016\cite{Symantec2016}. Blue Coat has devices to snoop encrypted internet traffic. Blue Coat became a SubCA under Symantec.  This unification increased the skepticism.
	
\end{itemize}

These fatal incidents lead to many researches to distribute the absolute trust on CAs to multiple authorities. To detect fake but valid TLS certificates, key pinning \cite{Langley2011,MP2012}, crowd sourcing \cite{Wendlandt:2008,AliKer2009,TR12015,EckerBurns2010,CertPat,Soghoian2012} , pushing revocation information to browsers \cite {Rivest1998,Langley2012} are the initial solutions which are partly implemented but failed due to scalability problems.

\subsection{Related Work} \label{relatedwork}
We here briefly describe the previous attempts for solving the issues described above in a chronological order and point out their potential weaknesses.

\subsubsection{Sovereign Key (SK) Cryptography}

\emph{SK} has been proposed by Eckersley in order to prevent  \emph{man-in-the-middle (MITM)} and \emph{server impersonation} attacks against domains protected by TLS certificates \cite{Eckersley2012}. In SK, domains generate a sovereign asymmetric key pair for a set of selected services such as https, smtps, and imaps and publish public part of the key in TimeLine Servers ($\mathcal{TS}$) along with their domain name. $\mathcal{TS}$ store entries in read and append only data structures. During the TLS handshake, domains generate a fake certificate by their sovereign key and append to the certificate chain. If the clients cannot verify the signature on this fake certificate with the associated public key in $\mathcal{TS}$, they refuse to make the connection to the service. However, $\mathcal{TS}$ does not return any verifiable proof to domains whether the sovereign key is appended to the log \cite{YuCR14}. Furthermore, another unrealistic strong assumption of \emph{SK} is that clients have to trust not only the $\mathcal{TS}$ but also their Mirrors \cite{YuCR14}.

\subsubsection{Certificate Transparency (CT)}

\emph{CT} is proposed by Google in 2013 and aims to detect fake but valid certificates by providing append-only, publicly auditable logs for all issued TLS certificates, and shorten their lifetime \cite {RFC6962}. \emph{CT} improves the \emph{SK}'s append-only log approach by using append-only Merkle Hash Trees (MHT) for appending and efficiently verifying TLS certificates \cite{crosby2009efficient}. In this respect, \emph{CT} introduces \emph{Certificate Log}s ($\mathcal{CL}$), \emph{Monitor}s, and \emph{Auditor}s as new entities. CAs submit the new TLS (pre)certificates to the $\mathcal{CL}$s which generates a cryptographic proof called Signed Certificate Timestamp (SCT). After appending the new certificates to the log, $\mathcal{CL}$  computes the Signed Tree Head (STH) and generates \emph{Consistency Proof} for proving whether the new log is an extension of the old one. \emph{Merkle Audit Proof} shows the existence of a certificate in the log. SCT can be delivered to the TLS clients either as a certificate extension, or in a TLS extension during TLS handshake, or in the OCSP response.  If the TLS clients do not receive SCT by  aforementioned-means or cannot verify them using $\mathcal{CL}$'s public key, they may refuse to connect to the service. Domain owners or private companies can act as \emph{Monitor}s and continually inspect certificates of their interest in $\mathcal{CL}$ whether there exist illegitimate certificates. Browsers, or in general TLS clients, act as \emph{Auditor}s. They verify if a $\mathcal{CL}$ is behaving properly and cryptographically consistent.


However, the authors in \cite{mazieres2002building} show that if an adversary can get a fake but valid TLS certificate and can control the $\mathcal{CL}$ then he can perform an impersonation/MITM attack to the targeted victims by providing a fraudulent view of the log that contains the fake certificate. This attack is later referred to as \emph{split-world attack} \cite{chuat2015efficient} which is elaborated as follows. 

\begin{enumerate}
	\item The adversary gets a fake but valid TLS certificate for a domain from the CA.
	\item The adversary or the CA submits this certificate to the log. 
	\item The adversary obtains a bogus SCT for the fake TLS certificate from the log. SCT is fake, since the log maintains more than one hash trees. It shows different views of the tree to different sets of clients. \emph{Merkle Audit Proof} and \emph{Consistency Proof}  are generated from different hash trees and the victim clients cannot understand that the proofs are not generated from the genuine hash tree.
	\item When a targeted victim client tries to connect to a domain, the adversary controlling the traffic sends the fake certificate and the bogus SCT to the client.
	\item Upon validation of the fake certificate and the SCT, client connects to either the fake domain or the real domain through a proxy.
\end{enumerate}

\emph{Monitor}s cannot detect this attack since they get and verify the \emph{Consistency Proof}s only from the genuine view of the log.  \emph{Auditor}s (victim TLS Clients) cannot detect the attack, since they can verify the existence of the fake TLS certificate in the log with the fake STH and the \emph{Merkle Audit Proof} generated from the fraudulent view of the log. The other \emph{Auditor}s also cannot detect either, since they get STH and the proofs from only the genuine view of the log. 

As pointed out in RFC 6962 for CT \cite{RFC6962}, to detect the attack in the existing \emph{CT} architecture,   there must be sufficiently large number of clients and servers gossiping their view of STHs with each other. Moreover, some of the clients should also act as \emph{Auditor}s for checking the consistency of the log. Some of these gossiping \emph{Auditor}s should be able to receive both the genuine and the fake proofs belonging to a log to figure out the inconsistency. If an \emph{Auditor} has two STHs with the same tree size but with different values, this will be an evidence for the misbehaviour of the log. If an \emph{Auditor}  has two different STHs with unequal tree sizes then it can request the \emph{Consistency Proof} from the log. Since the log cannot provide this proof it should be investigated for malicious behaviour.

\subsubsection{Certificate Issuance and Revocation Transparency (CIRT)}
\emph{CIRT} focuses on extending \emph{CT} to provide additional proofs so that CAs can show their honest behaviour\cite{Ryan14en}. Therefore, \emph{CIRT} maintains two merkle trees as public logs which are ordered chronologically and lexicographically. In addition to \emph{CT}'s proofs, \emph{CIRT} provides proofs for  whether a certificate is marked as revoked in the log, whether a certificate is current(not replaced and not revoked) and finally whether a certificate is absent(has never been issued).  
However, \emph{CIRT} is also vulnerable to the split-world attacks \cite{chuat2015efficient} as \emph{CT}.

\subsubsection{Accountable Key Infrastructure (AKI)}
\emph{AKI}  proposes a new PKI architecture for reducing the level of trust to the CAs where all the operations defined in a conventional PKI are performed with participation of more than one entity \cite {Hyn2013}. All entities either monitor or report the operations performed by other entities in order to distribute the accountability among the participating entities. \emph{Certification Agency (CA)}, \emph{Integrity Log Server (ILS) Operators (ILSO)}, and \emph{Validator}s are the new entities introduced by \emph{AKI}. \emph{CA}s still issue certificates to domains but they are not the absolute authority for certificate management anymore. \emph{ILSO}s store not only the certificates, but also their registration, update, and revocation information lexicographically according to domain names. \emph{Validator}s monitor \emph{ILS} operations and check whether they operate as expected. Domain owners can select the \emph{CA}s and the \emph{ILSO}s they trust,  define the minimum number of recommended \emph{CA} signatures on a certificate and the rules for certificate management. They register the certificate to one or more \emph{ILS}s. Domains send their certificates and the verification information received from the \emph{ILS}s to the clients. Clients verify the certificates using preinstalled trusted \emph{CA} certificates and the \emph{ILS} public keys.


However,  \emph{AKI}  makes a strong assumption that the trusted entities (CAs, \emph{ILS}s, and \emph{Validator}s) do not collude with each other which is unrealistic in case of a strong adversary who is willing to intercept the traffic.  A compromised CA and an \emph{ILS} is sufficient to generate a fake certificate in \emph{AKI}. The adversary gets proofs from the compromised entities in order to send to the TLS client. Taking this fact into consideration, it is also possible to make a split-world attack to the \emph{AKI}. Unfortunately, there is no way of detecting this attack in \emph{AKI}.

Secondly, certificate revocation is a weak point in \emph{AKI} since the domain owner can request the certificate revocation from an \emph{ILS} without requiring confirmation of any other parties. Namely, an adversary, which has compromised the domain private key, can request the revocation of the corresponding certificate without further verification. 

Finally, clients have to trust not only  the \emph{CA}s and the \emph{ILS}s but also the \emph{Validator}s in \emph{AKI} which is a burden for them in terms of trusted entity management.

\subsubsection{Attack Resilient PKI (ARPKI)}

\emph{ARPKI}  is an improvement of \emph{AKI}, which offers a security guarantee against adversaries which can compromise even $n - 1$ trusted entities\cite{Basin:2014}.  For generating an \emph{ARPKI} certificate (called \emph{ARCert}), at least two \emph{CA}s and one \emph{ILS} are required. For the initial registration process, a domain owner selects the trusted entities, at minimum two \emph{CA}s and one \emph{ILS} (i.e., \emph{CA}$_1$, \emph{CA}$_2$, and \emph{ILS}$_1$). The domain owner designates one of the \emph{CA}s (e.g., \emph{CA}$_1$) for validating \emph{CA}$_2$ and \emph{ILS}$_1$ operations and serving as a messenger between these entities and himself. \emph{ILS}$_1$ takes the responsibility for ensuring synchronization of \emph{ARCert} among majority of \emph{ILS}s. Domains send the cryptographic proof signed by three trusted entities along with the \emph{ARCert} to the clients. Upon verification of the proofs, clients connect to the domain. 


However, \emph{ARPKI} is also subject to split-world attack if it the entities required to generate an \emph{ARPKI} certificate collude together. There is also no detection mechanism for this attack in \emph{ARPKI}  as \emph{AKI}. Secondly, designating an \emph{ILS} for making synchronization with other \emph{ILS}s may lead to a single point of failure in \emph{ARPKI}.

\subsubsection{NameCoin and CertCoin}

\emph{NameCoin} is a cryptocurrency forked from Bitcoin\cite{nakamoto2008bitcoin}, which is designed to act as a decentralized DNS for ".bit" adresses \cite{NameCoin}. In \emph{NameCoin},  self signed TLS certificate of a domain can be added to the DNS addresses as auxiliary information. TLS clients can then authenticate the domain during TLS handshake using this certificate.

\emph{Decentralized Public Key Infrastructure with Identity Retention (Certcoin)} proposes a decentralized PKI architecture based on \emph{NameCoin} where no CA exists \cite{fromknecht2014certcoin}. In \emph{Certcoin}, the basic PKI operations are defined such as registering an identity with a public key, looking up, verifying, and revoking a public key for a given identity. Identities register an online and an offline key pair to themselves. Online key is used for domain authentication whereas offline key is used to revoke old online keys and to sign new online keys. However, in both proposals (i.e., Namecoin and Certcoin), there is no identity verification. Namely, whoever first claims the ownership of an identity owns it. Consequently, in real world, identities (in particular TLS clients) can easily be deceived.  Secondly, they have also no adequate solution in case both online and offline keys are compromised. Hence, the identity owners cannot reclaim their identities securely which can lead to unusable identities. Thirdly, since both proposals are using the Bitcoin blockchain architecture, verifying the owner of a public key, and looking up the public key of an identity is extremely inefficient. In order to solve these issues,  \emph{Certcoin} proposes extra entities which are also maintained in the blockchain such as accumulators and distributed hash tables. Since these entities are not a part of the blockchain architecture,  they can cause new complexities in terms of maintenance, authentication, and verification.

\subsubsection{Distributed Transparent Key Infrastructure (DTKI)} \label{sec:dtki}

\emph{DTKI} proposes a public log based certificate management architecture which minimizes oligopoly, prevents use of fake certificates, and claims being  secure even if all service providers collude all together \cite{YuCR14}. \emph{Certificate Log Maintainer ($\mathcal{CLM}$)} and  \emph{Mapping Log Maintainer} ($\mathcal{MLM}$) are the two new entities introduced by \emph{DTKI}. $\mathcal{CLM}$s keep all valid, revoked and expired certificates for a set of domains and provide proofs for existence or absence of them. $\mathcal{MLM}$ maintains the association between a set of domain names and the  $\mathcal{CLM}$s which are maintaining the logs for them. Mirrors maintain a full copy of the data stored by both the $\mathcal{CLM}$s and the $\mathcal{MLM}$. \emph{CA}s make identity checks and issue certificates, but they are not the sole entity for providing trust to connect to a domain.  Inspiring "soverign key" concept in \emph{SK}, a domain here owns two types of certificates, TLS certificate and a master certificate which is used for requesting a new TLS certificate from the \emph{CA}, and registering it to the $\mathcal{CLM}$. Users or in particular  browsers first query the $\mathcal{MLM}$ in order to find the correct $\mathcal{CLM}$ for a specific domain. To make a connection decision, first the proofs received from the  $\mathcal{MLM}$ is verified, then  $\mathcal{CLM}$  is queried in order to retrieve proofs for the domain's TLS certificate.



In \emph{DTKI}, assumes all master certificates are genuine, and fake master certificate issuance does not likely occur since \emph{CA}s are running businesses which cannot afford loss of reputation. However, this is not a valid argument since most of the fake certificates are generated due to lack of adequate security controls or processes. Namely, if the \emph{CA} and the $\mathcal{CLM}$ are both compromised, \emph{DTKI} would not be able to prevent fake Master and TLS certificate issuance.  From this perspective, the adversary who is controlling  $\mathcal{CLM}$ and capable of generating fake but valid Master and TLS certificates can make split-world attack to the targeted victims. This attack unfortunately cannot be detected because there is no monitoring process in \emph{DTKI} due to the assumption of genuine master certificates. 


\subsubsection {Blockchain-based Certificate and Revocation Transparency} \label{sec:bbctrt}

Very recently in FC'18, Wang et al. proposed a blockchain-based certificate and revocation transparency to store the TLS certificates and their revocation status (i.e., CRL and OCSP) \cite{WLCWJZ18}. Briefly in this scheme, web servers publish their TLS certificates to the blockchain using their \emph{publishing key pair}s which are used to sign the transactions. Note that these publishing key pairs are different from the key pair in the certificate and are initially certified by a certain set of web servers which already exist in the blockchain. In this scheme, the transactions have a validity period,  therefore TLS certificates and their revocation status are added to the blockchain periodically during their lifetime. During a TLS handshake, a web server sends a certificate transaction and its Merkle audit path to a TLS client which verifies its validity through its locally stored syncronized block headers. 

However, this proposal has the following drawbacks. 

\begin{itemize}
	
	\item It has an unreliable basis for providing the trustworthiness of \emph{publishing key pair}s. Namely, a strong adversary, who can get fake but valid TLS certificates from corrupted CAs, can create some bogus domains (i.e., web servers) in advance and can use them to generate a valid signature of a \emph{publishing key pair} transaction. This problem occurs due to the trust to the web servers. The authors propose to solve this issue by having "more publicly-trusted CA"s to invalidate the interfering transactions. However, this introduces a trust level issue which is not explicitly clarified. 
	
	\item  For the revocation transparency, it relies on the CAs to publish the revocation data of the TLS certificates on the blockchain. However, the compromised or malfunctioning CAs may not issue CRL or  give OCSP response to the client in the specified time. 
	
	\item It is subject to MITM attacks where an adversary can convince a client with an unexpired transaction of a revoked TLS certificate. More concretely, during a TLS handshake, web servers send a certificate transaction to a TLS client to validate the TLS certificate. The TLS client accepts this transaction if it is not expired and is added to a confirmed block. However, a revoked or updated TLS certificate can also have an unexpired certificate transaction in the blockchain. Therefore, once an adversary sends this unexpired certificate transaction with its Merkle proofs to a TLS client, it is accepted during the TLS handshake. The TLS clients cannot detect the final state of the certificate since the clients only check the existence of the transaction in the corresponding block.
	
	\item The proposal is also inefficient in terms of storage costs due to following design considerations. 1) A TLS certificate is added to the blockchain periodically during its lifetime, 2) A CRL can be added to the blockchain for each revoked certificate (i.e., the number of CRL insertion to the blockchain is equal to the number of revoked certificates), 3) \emph{Publishing key pair}s are added to the blockchain periodically 4) It has large size headers which comprise DNS names existing in the transactions of the block.	
	
\end{itemize}

\subsection{Ongoing Security Issues}  \label{problem}

In the afore-mentioned public log based proposals, a strong adversary who has the ability to control the trusted entities (i.e., CA, Log Operator) can apply split-world attacks by providing different views of the logs to the targeted victims \cite{mazieres2002building}. While some of these proposals cannot detect this attack, others propose to use gossip protocols to identify it by ensuring a consistent view of the log for the TLS clients \cite{chuat2015efficient, gossip1, gossip2,  7546521}. Still, this attack can only be identified if 1) there are sufficient numbers of gossiping TLS clients and servers, 2) at least some of them are able to view the genuine log and request the consistency proof from the log.

As also described in \cite{liu2015end},  certificate revocation and validation processes have major problems in today's PKI.  Namely, for the revocation process, certificate owners have to rely on CAs which have the full responsibility to revoke the certificates and give proper revocation services. However, a compromised or malfunctioning CA may not behave as expected. Browsers would then accept revoked certificates since they rarely check whether the certificates are revoked\cite{liu2015end}. 
Moreover,  incompatible and inadequate implementations of certificate validation and revocation behavior within browsers are also error prone \cite{liu2015end}.

Another issue is the necessity of trusted key management in TLS clients. TLS clients have to trust CA certificates or some other trusted entities' public keys  to make a successful TLS certificate validation. In case of a compromise,  removal of root certificates/keys from all the clients' trusted key stores brings burden (e.g., due to IT policy restrictions, OS configuration, network communication) and causes vulnerability (e.g., outdated OS or apps). Moreover, if an adversary can inject a fake CA certificate to the trusted certificate store of a client, he can easily perform MITM attack without being detected \cite{Lenovo2015,HelnetSecurity}.

\subsection{Our Contributions}  \label{contribution}
In this paper, we propose a new and efficient PKI architecture, which we called \emph{CertLedger}, to make TLS certificates and their revocation status transparent while eliminating the above-mentioned issues. \emph{CertLedger} uses a blockchain based public log to validate, store, and revoke TLS certificates. In summary, we make the  following contributions in \emph{CertLedger}:

\begin{itemize}
	\item Resistance to split-world attacks
	\item More transparent revocation process rather than relying on CAs
	\item Unique, efficient, and trustworthy certificate validation process
	\item Trusted key store elimination in the TLS clients
	\item Preserving the privacy of the TLS clients
	\item No external auditing requirement due to inherent public log architecture  
	\item Efficient and prompt monitoring to minimize the attack duration
	\item Transparency in trusted CA management  
\end{itemize}

We  also provide a detailed security and performance analysis of \emph{CertLedger} and compare it with the most recent works.

\subsection{Roadmap}
In Section \ref{whypermissionless}, we describe how public blockchain solves the afore-mentioned problems of PKI. In section \ref{CertLedger}, we describe our proposal \emph{CertLedger} which is a new PKI model with certificate transparency based on blockchain. We analyse \emph{CertLedger} in terms of security and performance  in Sections \ref{securityanalysis} and \ref{performance} respectively. In Section \ref{comparison}, we compare \emph{CertLedger}  with existing schemes and  finally conclude the paper with describing the future work in Section \ref{conclusion}.

\section{How Public Blockchain Solves PKI Problems?} \label{whypermissionless}

Blockchain is a shared, immutable, decentralized public ledger comprising an ever growing list of blocks. A block is a data structure which is comprised of a header and a list of transactions. Each block is linked to the previous one with a cryptographic hash, therefore blocks are inherently secured from tampering and revision. Blockchain network is a decentralized peer-to-peer (P2P) network composed of full nodes and light nodes. Full nodes store a copy of the blockchain, validate, and propagate new transactions and blocks across the network while light nodes store only block headers. All nodes can create transactions to change the state of the blockchain. New blocks of the transactions are collectively validated and appended to the existing chain according to a distributed consensus mechanism. There are several consensus mechanisms used in blockchain networks, e.g., Practical Byzantine Fault Tolerance algorithm (Practical BFT) \cite{castro1999practical} (which is utilized in HyperLedger Fabric \cite{HyperLedger}), Delegated BFT \cite{NeoWP} (which is utilized in Neo\cite{Neo}), Proof-of-Work (PoW) (which is utilized in Bitcoin \cite{nakamoto2008bitcoin}, Ethereum \cite{wood2014ethereum}), Ripple \cite{schwartz2014ripple} (which is utilized in Ripple\cite{Ripple}), and Proof-of-Stake (PoS) \cite{kiayias2017ouroboros} (which is utilized in Cardano \cite{Cardano}, PeerCoin \cite{king2012ppcoin}).  From now on, for simplicity, we we refer to 'blockchain' as blockchain network. In general there are also three types of Blockchain: 1) Permissionless (e.g., Bitcoin \cite{nakamoto2008bitcoin}, Ethereum \cite{wood2014ethereum}, ZCash \cite{sasson2014zerocash}), 2) Public Permissioned Blockchain (e.g., Ripple\cite{schwartz2014ripple}), 3) Private Permissioned Blockchain. While any peer can join and leave the network any time in permissionless blockchains, permissioned blockchains require authorization for the membership of the peers. In the permissioned blockchains,  if public verifiability is required, then the transactions of the blockchain must also be public. However, private permissioned blockchains may be used for corporate networks which have sensitive data.

In the following, we show why blockchain solves 1) Split-world attack, 2) Certificate revocation and validation problems, and 3) Trusted certificate/key store management problems.

 \begin{enumerate}
 	\item As described in Section \ref{intro},  existing proposals try to solve the transparency issue in PKI by introducing (one or more) public logs  \cite{RFC6962, Hyn2013, Basin:2014, YuCR14}. However, they are still subject to split-world attack due to the trust to a log operator. In order to prevent such an attack, the trust should be distributed in such a way that a single log operator could not be able to control the log itself. Therefore, a decentralized public log mechanism is required which is synchronously updated only upon a consensus of its clients. Once the consensus is achieved, the log should be updated and could not be reverted anymore. We would also like to highlight that, in order to prevent the split-world attack, RFC 6962 for CT states \cite{RFC6962}, "\emph{All clients should gossip with each other, exchanging STHs at least; this is all that is required to ensure that they all have a consistent view. The exact mechanism for gossip will be described in a separate document, but it is expected there will be a variety}".  
 	Therefore,  as also directly implied by "\emph{all clients gossiping to each other}",  blockchain is an architecture fullfilling all these required features.
 	

 	\item Certificate validation is a required process for each TLS connection, and comprises trusted path construction and revocation checking phases. Currently, certificate validation burden is entirely on browsers. However, if TLS certificates are stored on a blockchain, trusted path construction can be done only once while they are being appended to the blockchain. Consequently,  browsers can trust the TLS certificates on the blockchain without requiring further validation. As also described in \cite{liu2015end},  browsers have different implementations for revocation checking which are error prone. By managing the revocation status of TLS certificates in the blockchain, revocation checking process can be simplified and unified. Namely, revocation checking burden using CRL and OCSP services can be eliminated. Furthermore, while certificates can be revoked only by CAs in the existing system, they can also be revoked by their owners on the blockchain. Since, the complete revocation process is conducted through the blockchain it becomes more transparent.
 	\item Validation of certificates while appending to the blockchain also requires trusted CA certificates to exist on the blockchain. Therefore, these certificates have to be stored and managed on the blockchain. Consequently, TLS clients do not need to store trusted CA certificates anymore since blockchain ensures to append the TLS certificates issued only from the trusted CAs.
 	
 \end{enumerate}




\subsection{Blockchain Characteristics for PKI}

Using the decision-sequence of Wüst and Gervais \cite{wust2017you}, we now identify the type of blockchain to manage the certificate log. Note that, a \emph{writer} is an entity who is able to accumulate new transactions into a new block and append it to the blockchain.

\begin{enumerate}
	\item \emph{Do you need to store state? } \newline TLS certificates are being updated continually due to expiration or revocation. The state of the certificates have to be stored and updated whenever necessary.
	
	\item \emph{Are there multiple writers?} \newline TLS certificates generated by the trusted CAs are appended to the log. A manipulated single writer can append fake but valid certificates to the log,  delay or ignore appending the genuine ones. Therefore, increasing decentralization while writing to the log will reduce the risk of manipulation due to the fact that broad participation of \emph{writer}s will lead to a more reliable and robust log.  
	
	\item \emph{Can you use an always online Trusted Third Party (TTP)?} \newline As described in Section \ref{intro}, a strong adversary can control any TTP which can lead to a single point of failure.  An online TTP assumption is the main source of vulnerability.
	
	\item \emph{Are all writers known?} \newline The \emph{writers} may be known or unknown. However, if they are known, they should be selected and dispersed all over the world in such a way that their malicious cooperation and manipulation could not be possible.
	
	\item \emph{Are all writers trusted?} \newline Even though, all the \emph{writers} are seemingly trusted, some of them can be controlled by a strong adversary.  
	
	\item \emph{Is public verifiability required?} \newline The existence and validity status of all TLS certificates must be verifiable by public for ultimate transparency. 
	
\end{enumerate}

Thus, the decision flowchart results in either a permissionless or a public permissioned blockchain for managing the certificate logs. However, the blockchain requires the following additional features. First of all, it must comprise smart contract infrastructure to implement the required rules for validating the state transitions \cite{SmartContract}.  Secondly, the underlying consensus mechanism should not lead to temporary forks, since some of the TLS clients can verify an incorrect state of a TLS certificate before the blocks have been fully confirmed. Thirdly, the required time for the confirmation of a new block in the consensus mechanism should not be high, so that the transactions can change the state of the TLS certificates in an acceptable time frame. Finally, blockchain architecture should enable TLS clients to verify the final state of the TLS certificates and generate Merkle proofs efficiently. \footnote{State Merkle Patricia Trees are generally maintained to generate proofs and track the final states of the assets efficiently. The Merkle root of this tree is stored in the block headers, so that the integrity of the tree, and the state proofs generated out of it can be verified. \cite{PatriciaTree} } 

In this respect,   \emph{CertLedger}  can be deployed on an existing  blockchain architecture fulfilling these requirements such as Ethereum \cite{wood2014ethereum}, Neo\cite{Neo} and Ontology \cite{Ont}. In these architectures, any consensus mechanism can be selected which will not lead to temporary forks such as PBFT\cite{castro1999practical} and DBFT \cite{NeoWP}.

\section{CertLedger } \label{CertLedger}

\emph{CertLedger} is a PKI architecture to validate, store and revoke  TLS certificates and manage Trusted CA certificates on a public blockchain. It aims to make certificate issuance and revocation life cycle more transparent and to eliminate any kinds of MITM attacks. Moreover, it also aims to unify certificate validation process for all TLS clients due to its inconsistent and inadequate implementations throughout different TLS clients.

\emph{CertLedger} manages the PKI functionalities through \emph{state objects}. A state object is a digital document which is comprised of data and an immutable smart contract code to manage it. Each state object has a unique address in the blockchain. State changes of the assets are triggered by transactions and tracked through the state objects. \emph{CertLedger}  comprises the following state objects.

\begin{itemize}
	\item \emph{Domain State Object} stores and manages the states of all TLS certificates and  their revocation status. This state object comprise the necessary code for validating the TLS certificate according to international standards such as RFC 5280 \cite{rfc5280}. It uses the \emph{Trusted CAs State Object} while building a trusted path for the TLS certificate.  Moreover, it also comprises the necessary code for changing the status of the TLS certificate as 'revoked'.  Its smart contract validates a TLS certificate while adding to the \emph{CertLedger} with following sample flow.
	
	\begin{itemize}
		\item Check whether the certificate has already been added.
		\item Check whether the certificate is in its validity period.
		\item Check whether the certificate  is compliant to the TLS certificate profile.
		\item Check whether the certificate is signed by one of the CA certificates in  \emph{Trusted CAs State Object}.
		\item Store the  new TLS Certificate in the  \emph{Domain State Object} and set its revocation status as 'Not Revoked'.
	\end{itemize} 
	
	\item \emph{Trusted CAs State Object} stores the set of the trusted CA certificates and smart contract code for adding new CA certificates as 'trusted' or changing the status of the existing ones as 'untrusted'. The smart contract code makes all the necessary controls on the certificate and checks whether it is complaint to international standart certificate profiles (e.g. RFC 5280 \cite{rfc5280}) for being a CA certificate. 
	
	\item An \emph{Account State Object} stores the \emph{CertLedger} token balance of a \emph{CertLedger} entity and  is controlled by the account's private key(s).  This state object is used to create and trigger any transaction within \emph{CertLedger}.
	
	\item \emph{CertLedger Token State Object} is the source of the initial token supply. It comprises the smart contract code to determine the initial owner of the token supply, to transfer token between different \emph{Account State Object}s and to give permission to a state object for transfering a certain amount of token from a given \emph{Account State Object}.
		
	\item \emph{Fraud Report State Object} stores fraud reports about the CAs which are in the \emph{Trusted CAs State Object}. The reports comprise proofs for the fraudulency of the CAs which but possibly issued a valid but fake certificate. However, the accused CAs can also add proofs about their trustworthiness.
	
\end{itemize}

\subsection{Entities}
We have three types of entities in our model: 1) CertLedger Entities (\emph{CertLedger Board}, \emph{CertLedger Foundation}, \emph{CertLedger Client}s ) 2) External Entities (\emph{Certificate Authorities (CA)}, \emph{Domain Owner}s) 3) Underlying Blockchain Entities (\emph{Miner}s, and \emph{Full Node}s).

\begin{itemize}
	\item A \emph{CA} has basically four different tasks: 1) Checks  the identity of the \emph{Domain Owner} for his TLS certificate request. 2) Issues a TLS certificate upon succesfull identity verification. 3) Optionally, creates a transaction to append the issued certificate to the \emph{CertLedger}. 4) Optionally, creates a transaction to change the revocation status of the certificate in the \emph{CertLedger}. Therefore, in our model, \emph{CA}s  do not issue Certificate Revocation List (CRL) and give OCSP service anymore. 
	
	\item \emph{CertLedger Board} is a trusted organization who manages \emph{Trusted CAs State Object}. It defines the standards and procedures to manage this state objects. In order to distribute the trust, private key of its \emph{Account State Object} is shared among the \emph{CertLedger Board Member}s in a threshold fashion, i.e., for $n$ board members at most $t$ of them are assumed to be corrupted. 
	
	In order to provide transparency, the requirements of being a board member should be defined as an international standard by the organizations such as IEEE\cite{IEEE}, ISO\cite{ISO} and IETF \cite{IETF}. Certificate Authorities, Browser and O/S Development Companies(Foundations), Research Institutes, Universities  and International Standardization Organizations (ETSI\cite{ETSI}, ISO\cite{ISO}, IETF \cite{IETF} etc.) are the natural candidates of the \emph{CertLedger Board}, but any organization in the world fulling the requirements can be a board member. 
	
	\item \emph{CertLedger Foundation} promotes, supports and develops CertLedger platform and does research activities. They are also the owner of the initial \emph{CertLedger} token supply.
	
	\item A \emph{CertLedger Client} is a light node(of the underlying blockchain) which stores only the block headers. It has connections with its peers according to the underlying blockchain light client protocol. It verifies the TLS certificates of the domains through \emph{CertLedger}. 
	
	\item A \emph{Miner} is a \emph{writer}, a full node  which selects pending transactions from the pool, validates them, and then creates new blocks according to the consensus protocol.  It generates blocks for all the transactions of the underlying blockchain.
	
	\item  A \emph{Full Node} generates proof for the \emph{CertLedger Client}s to verify the final state of the state objects. 
	
	\item A \emph{Domain Owner} offers secure services to the \emph{CertLedger Client}s through protocols such as https, imaps, and sips. He has the following tasks: 1) Makes a TLS certificate request to \emph{CA}s for his domain. 2) Optionally,  creates a transaction to append the received TLS certificate to the \emph{CertLedger}.  3) Monitors his up-to-date TLS certificate in the \emph{CertLedger} (e.g., using event watchers on Ethereum). 4) In case of compromise detection, immediately creates a transaction to revoke his TLS certificate and creates another transaction to report the fraud.
	


\end{itemize}

\subsection{Trust and Threat Model} \label{trustthreatmodel}

\subsubsection{Adversary Capabilities} \label{adversary}

\emph{Certificate Authorities (CA)},  \emph{CertLedger Client}s, and \emph{CertLedger Miner}s are assumed to be malicious which can behave arbitrarily, e.g., fake but valid certificates can be issued from the CAs which may be either corrupted or operated with inadequate security policies. Furthermore, for $n$ \emph{CertLedger Board Member}s at most $t$ of them are assumed to be corrupted.

\subsubsection{Assumptions}
We first assume that the network underlying \emph{CertLedger} is insecure and a certain number of \emph{CertLedger Client}s are honest with respect to the underlying consensus model for agreeing on the honest state of the blockchain.

Secondly, we assume that the underlying cryptographic primitives and the blockchain architecture of the \emph{CertLedger} are secure. 

\subsection{PKI Functionalities of CertLedger}\label{architecture}

\begin{figure*}
		\begin{center}
			\scalebox{0.33}{\includegraphics{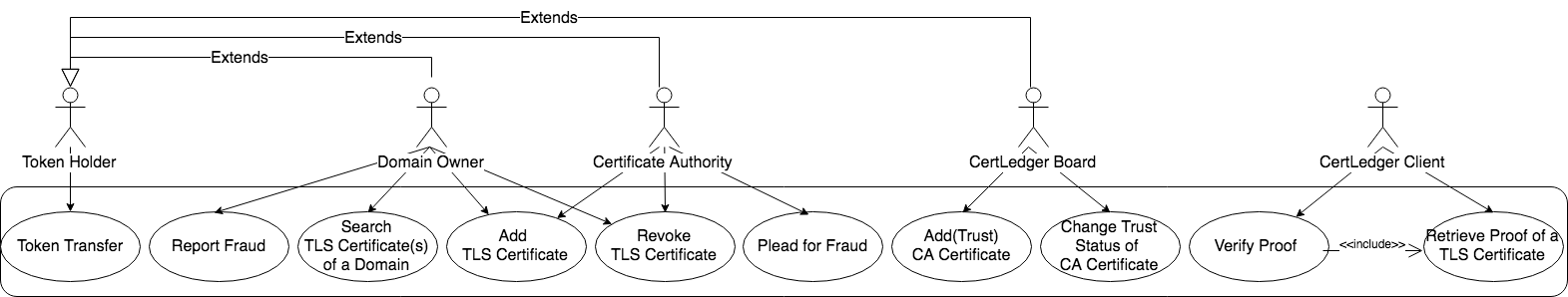}}	
		\end{center}
		\captionsetup{justification=centering}
		\caption{Functionalities of CertLedger}
		\label{UseCase0}
	\end{figure*}

We now describe the functionalities of the \emph{CertLedger}. The responsibility of the entities in these stages are illustrated in Figure \ref{UseCase0}.

\begin{itemize}
	\item \textbf{Adding a new Trusted CA Certificate}: \emph{CertLedger Board}  adds  a CA certificate to the \emph{Trusted CAs State Object} as follows.
	
	\begin{itemize}
		\item Upon application of a CA,	\emph{CertLedger Board} audits the CA and verifies whether it complies with the Trusted CA standards. 
		\item Upon a successful audit, they generate a transaction comprising the CA certificate which is signed in a threshold fashion to add the CA certificate to the \emph{Trusted CAs State Object}.
		\item Smart contract code in the  \emph{Trusted CAs State Object} verifies the signature of the transaction, and makes all the necessary checks on the CA Certificate.
		\item  Upon successfull validation of the CA certificate it is added to the \emph{Trusted CAs State Object} as a trusted certificate.
		\item \emph{Trusted CAs State Object} triggers \emph{CertLedger Token State Object} to transfer the operation fee  from \emph{CA}'s \emph{Account State Object} to the \emph{CertLedger Board}'s \emph{Account State Object}. 
	\end{itemize}
	
	\item \textbf{Changing Trust Status of a CA Certificate}: If a trusted \emph{CA} fails to comply to Trusted CA standards in further audits or its misbehavior is proved through a fraud report, then the status of its certificate is set as 'untrusted' in the  \emph{Trusted CAs State Object} as follows.
	
	\begin{itemize}
	\item \emph{CertLedger Board Member}s generate a transaction which is signed in a threshold fashion to change the  status of the CA certificate in the \emph{Trusted CAs State Object}.
	\item Smart contract code in the  \emph{Trusted CAs State Object} verifies the signature of the transaction, and checks whether the CA certificate is unexpired and exists in the \emph{Trusted CAs State Object}.
	\item The state of the CA certificate is set as 'untrusted' in the \emph{Trusted CAs State Object}.
	\item \emph{Trusted CAs State Object} triggers \emph{Domain State Object} to change the state of all TLS certificates issued from this CA  as 'revoked'.
	\item \emph{Trusted CAs State Object} triggers \emph{CertLedger Token State Object} to transfer the operation fee  from \emph{CA}'s \emph{Account State Object} to the \emph{CertLedger Board}'s \emph{Account State Object}. 
	
\end{itemize}	
	\item \textbf{Adding a new TLS certificate}: As in the conventional PKI architecture, upon application of a \emph{Domain Owner}, \emph{CA} performs the relevant verifications according to its policy and issues the TLS certificate. Note that a \emph{Domain Owner} can have more than one valid TLS certificate of a domain.  In practice, before the expiration of his TLS certificate, the \emph{Domain Owner} receives a new TLS certificate from the \emph{CA} with overlapping dates, so that the latter becomes active before the expiration of the former. 
	
	Upon generation of a new TLS certificate, \emph{Domain Owner}s and \emph{CA}s add it to the \emph{Domain State Object} as follows.

	 \begin{itemize}
	 	\item \emph{Domain Owner}/\emph{CA} creates a new transaction comprising the new TLS certificate and signs it using the private key of his own \emph{Account State Object}.
	 	\item Smart contract code in the \emph{Domain State Object} validates the certificate if the \emph{Domain Owner}/\emph{CA}'s balance is sufficient for invoking the transaction. Upon a successful validation, the new TLS certificate is added to the \emph{Domain State Object}.
		\item \emph{Domain State Object} triggers \emph{CertLedger Token State Object} to transfer the operation fee  from \emph{Domain Owner}/\emph{CA}'s \emph{Account State Object} to the \emph{CertLedger Foundation}'s \emph{Account State Object}. 
	 	\item \emph{Domain State Object} triggers an event notification after addition of the new certificate.
	 \end{itemize}

	\item \textbf{Revocation of a TLS Certificate}: 
	
%
%
	
	A TLS certificate is revoked in the \emph{CertLedger} as follows (Fig. \ref{FigRevocationUpdate}):
	
	\begin{itemize}
		
		\item  A valid revocation request can only be generated for the non-expired and valid TLS certificates. In our model, \emph{Domain Owner}s and \emph{CA}s are assumed to be the only parties creating a valid revocation request.  The revocation request of a certificate comprises a signature generated by either itself or by its issuing \emph{CA}'s certificate. The transaction comprising the revocation request is signed by the private key of the related entity's \emph{Account State Object}. 
		
		\item Smart contract code in the \emph{Domain State Object} validates the transaction if the user's balance is sufficient for invoking the transaction. If the revocation code succeeds, then the state of the certificate is changed as 'revoked' in the \emph{Domain State Object}.
	
		\item \emph{Domain State Object} triggers \emph{CertLedger Token State Object} to transfer the operation fee  from \emph{Domain Owner}/\emph{CA}'s \emph{Account State Object} to the \emph{CertLedger Foundation}'s \emph{Account State Object}.
		
		\item \emph{Domain State Object} triggers an event notification after revocation of the certificate.

	\end{itemize}

	\item \textbf{Reporting a Fraud}: The fraud reports must also be transparent, and should not be submitted and processed in conventional mechanisms (e.g., email or personal applications). If the \emph{Domain Owner} detects a fake but valid TLS certificate for his domain in the \emph{CertLedger}, he reports the fraud in \emph{CertLedger} as follows.
	
	\begin{itemize}
		\item \emph{Domain Owner} generates  a transaction comprising the fake but valid TLS certificate and a signature generated with his genuine TLS certificate of the domain for proving his ownership to the domain.
		\item He signs the transaction with the private key of his \emph{Account State Object} and triggers the smart contract code of the \emph{Fraud Report State Object}.
		\item In the smart contract code, the existence of the fake but valid TLS certificate in the \emph{CertLedger} is verified, subject alternative names in the genuine and the fake TLS certificates are cross-checked, and the genuinity of the TLS certificate used for generating the signature in the transaction is verified. 
		\item If the contract code succeeds, the fraud report is added to the \emph{Fraud Report State Object}.
		\item The \emph{Fraud Report State Object} triggers the \emph{CertLedger Token State Object} to transfer the operation fee  from \emph{Domain Owner}'s \emph{Account State Object} to the \emph{CertLedger Board}'s \emph{Account State Object}.
		\item \emph{Fraud Report State Object} triggers an event notification after adding the new fraud report.
	\end{itemize}

	\item \textbf{Pleading against a Fraud}: 
	\emph{CertLedger Board} and the \emph{Trusted CA}s continually monitor the \emph{Fraud Report State Object} through event watchers. 	A \emph{CA} can plead in the \emph{CertLedger}, when it catches a notification triggered from the \emph{Fraud Report State Object} which comprises a fraud charge against it.

	\begin{itemize}
		\item Upon emergence of a new fraud report, it  immediately puts all the necessary documents for the issuance of the TLS certificate to a portal of the  \emph{CertLedger Board}.
		\item Generates a transaction comprising the hash of these documents signed with its CA certificate.
		\item Signs the transaction with the private key of his \emph{Account State Object} and triggers the smart contract code of the \emph{Fraud Report State Object}.
		\item Smart contract code verifies whether the plea is signed by the issuer of the fake TLS certificate and the CA certificate exists in the \emph{Trusted CAs State Object}.
		\item The \emph{Fraud Report State Object} triggers the \emph{CertLedger Token State Object} to transfer the operation fee  from \emph{Domain Owner}'s \emph{Account State Object} to the \emph{CertLedger Board}'s \emph{Account State Object}.
	\item \emph{Fraud Report State Object} triggers an event notification after adding the new fraud report.
		\item An event notification is triggered after adding the new plea against the fraud.
	\end{itemize}	

	Upon generation of the new block comprising this transaction, \emph{CertLedger Board} makes a decision whether there exists a fraud. If the \emph{CertLedger Board} is not convinced by the plead, then it creates a transaction to change the status of this CA certificate as 'untrusted'. 

	\item \textbf{Transferring Token}: Any entity having an \emph{Account State Object} can transfer token ownership. While entities transfer token to be able to trigger some of the functionalities of \emph{CertLedger}, they can also transfer token for only trading purposes. Token ownership is transferred as follows.

	\begin{itemize}
		\item An entity creates a transaction comprising the amount of the token to be transferred to a recipient account and the destination \emph{Account State Object} address.
		\item He signs the transaction with the private key of his \emph{Account State Object}.
		\item The transaction triggers the \emph{CertLedger Token State Object}.
		\item Smart contract verifies the signature of the transaction, and whether the balance of the entity is sufficient to make the transfer, and transfers the token from the sender's account to the recipient's account.
	\end{itemize}

	\item \textbf{Searching TLS Certificate(s) of a Domain}: Any entity of the \emph{CertLedger} can search the TLS certificate(s) of a domain existing in the \emph{CertLedger} (possibly with some parameters) to check their state. If the \emph{Domain Owner} finds a fake but valid TLS certificate, he immediately reports a fraud. 
	
	\item \textbf{Retrieving a State Proof of a TLS Certificate}: \emph{CertLedger Client}s and \emph{Domain Owner}s query any of the \emph{Full Node}s to receive proof for the existence and state of a TLS certificate. \emph{Full Node} returns the merkle proof generated out of the state Merkle Tree.  

	\item \textbf{Verifying a State Proof of a TLS Certificate}: \emph{CertLedger Client}s check whether the proof is generated for the requested TLS certificate, and verify it using the state Merkle Tree hash existing in the related block header.
	
	\item \textbf{Secure Communication through TLS:} 
	During a TLS handshake, a \emph{CertLedger Client} and a domain agree upon the latest block number. The domain sends the TLS certificate with its proof (generated out of the State Merkle Tree for the agreed block) to the \emph{CertLedger Client} through a TLS extension. The \emph{CertLedger Client} checks whether the TLS certificate is issued for the domain, and is in its validity period. It verifies the proof using the state tree hash which exists in the agreed block header. It also checks whether the proof is indeed generated for the  certificate and the state of the certificate is 'valid'. We highlight here that \emph{CertLedger Client}s do not require to make further validation of the TLS certificate, since this process is already completed while appending certificate to the blockchain.  
	
	\begin{figure}
		\begin{center}
			\scalebox{0.33}{\includegraphics{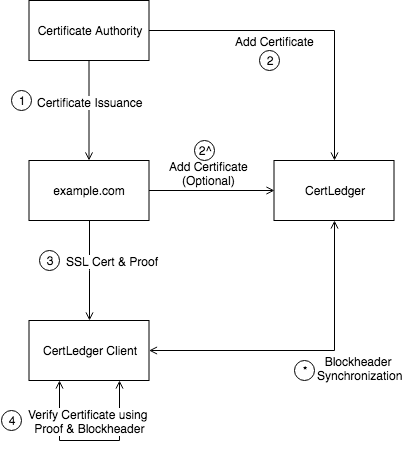}}	
		\end{center}
		\captionsetup{justification=centering}
		\caption{TLS Connection within CertLedger}
		\label{TlsConnection}
	\end{figure}

\end{itemize}

\begin{table*}
	\caption{Security comparison of Log Based Approaches to Certificate Management}
	
	\label{table:securityanalysis}
	\footnotesize
	\scalebox{0.92}{
		\begin{minipage}{17cm}
			\begin{tabular}{|l|c|c|c|c|c|c|}
				\cline{2-7} 
				\multicolumn{1}{c|}{} & \textbf{CT} \cite {RFC6962}& \textbf{AKI}  \cite {Hyn2013} & \textbf{ARPKI} \cite{Basin:2014}& \textbf{DTKI} \cite{YuCR14} & \cite{WLCWJZ18} &\textbf{CertLedger (This paper)} \tabularnewline
				\hline
				\makecell[l]{Resilient to split-world/MITM attack} & No & No & No  & No &  No &Yes \tabularnewline
				\hline
				\makecell[l]{Provides revocation transparency}  & No & Yes & Yes  & Yes & Partly\footnote{Relies on CAs and the revocation process is not transparent} &Yes \tabularnewline
				\hline
				\makecell[l]{Eliminates client certificate validation process} & No & No  & No & No  & Partly\footnote{Root certificate validation is still necessary} &Yes \tabularnewline
				\hline
				\makecell[l]{Eliminates trusted key management} & No & No  & No &  No  & No &Yes \tabularnewline
				\hline
				\makecell[l]{Preserves client privacy}  & No & Yes & Yes & No &  Yes &Yes  \tabularnewline
				\hline
				\makecell[l]{Require external auditing}  & Yes & Yes & Yes&  Yes  &  No &No \tabularnewline
				\hline 
				\makecell[l]{Monitoring promptness} & No & No & No &  No \footnote{Do not make monitoring due to assuming all master certificates are genuine}  & No \footnote{Blockchain architecture is inadequate} &Yes \tabularnewline
				
				\hline
			\end{tabular}
			
		\end{minipage}
		
	}	
\end{table*}

\section{Security Analysis} \label{securityanalysis}

We next analyse the security, usability, and privacy of \emph{CertLedger} according to the following properties. We also compare \emph{CertLedger} with the existing proposals from security and privacy perspective in Table \ref{table:securityanalysis}. \newline

\textbf{Property 1:} \emph{Under the assumptions of at most $t$ out of $n$ board members are corrupted and the underlying consensus protocol of the blockchain is secure, CertLedger eliminates the split-world attacks.}

Split-world attacks are only applicable if an attacker can present different views of the logs to the targeted victims. This can only exist when there is a trusted log operator and targeted victims do not exchange their view of the logs through gossip protocols. \emph{CertLedger} eliminates this attack since there is no trusted log operator and all of  the clients have the same up-to-date log. Namely, there is only one single log in the \emph{CertLedger} which can be updated only with the consensus of its clients. \emph{CertLedger}'s trustworthiness is inherently based on its underlying consensus mechanism. Since \emph{CertLedger Client}s verify the TLS certificate according to the view of the log formed after either the latest fully confirmed block or the latest block with respect to the underlying consensus mechanism, \emph{CertLedger Client}s are not effected from the convergence of the ledger. Under the assumption of the underlying consensus protocol of the blockchain architecture is secure, all \emph{CertLedger Client}s will have the final state of the TLS certificates. They are not subject to split-world attacks.

If more than $t$ board members are corrupted, a fraudulent CA can be added to the \emph{Trusted CAs State Object}. With a fake TLS certificate issued by this CA, split-world attack cannot be performed only to a selected set of victims. This is because the rest of the TLS clients would not be able to verify the genuine TLS certificate and make a successful TLS handshake. Moreover, the fake certificate will be promptly detected by the \emph{Domain Owner}s. Therefore, not only the fraudulent CA, but also the fraudulent board members would be responsible for this attack. Consequently, the community lose trust to the \emph{CertLedger} which results in either replacement of these board members or creation a fork of \emph{CertLedger} with a new set of board members.
\newline

\textbf{Property 2:} \emph{Revocation process of  CertLedger reduces depency to \emph{CA}s while making revocation status of all the TLS certificates transparent.}

In the \emph{CertLedger}, revocation process is not totally under the control of the \emph{CA}s. A valid revocation transaction will change the status of the certificate when it is included in a new block. The security of revocation transparency in \emph{CertLedger} is analysed as follows:

\begin{itemize}
	\item If the TLS certificate is not compromised, but private key of the TLS certificate is unusable due to forgotten password, lost or malformed private key, then the issuing \emph{CA} of the TLS certificate can create a revocation transaction. However, in this case  revocation of the TLS certificate is not necessary and if the private key exists,  the  \emph{Domain Owner} can simply destruct it.

  	\item If the TLS certificate is compromised, and if the  \emph{Domain Owner} can use the private key this TLS certificate, he can create a revocation transaction anytime without requiring the consent of the \emph{CA}. 
	
	With a compromised valid TLS certificate,  an adversary can also create a revocation request for this certificate. The adversary can have the following motivations with a compromised valid TLS certificate. 1. To make impersonation/eavesdropping attack, 2.  To interrupt the service of the domain. If the adversary has the first motivation, he does not create a transaction to revoke the certificate. If his motivation is the latter, then he creates a transaction to revoke the compromised TLS certificate of the domain. However, as soon as the \emph{Domain Owner} becomes aware of the compromise, his expected behaviour would also be to revoke the compromised TLS certificate immediately and continue the service using a different TLS certificate. Hence, revocation of the compromised TLS certificate by an adversary is not a vulnerability, but a prefered behaviour by the \emph{Domain Owner}.
	
	
	If the \emph{Domain Owner} cannot use the private key of the compromised TLS certificate,  he can not use this TLS certificate to secure his service and has to use a different TLS certificate of the domain. However, if the compromised TLS certificate is not revoked and the adversary has its private key, then he can perform an impersonation attack  if he can also control the DNS servers. Consequently,  in this case, \emph{Domain Owner} cannot create a transaction for revocation of the TLS certificate and requests it from the \emph{CA}. 
\end{itemize}

Hence, in \emph{CertLedger}, \emph{CA}s are not the only authority who can revoke a certificate. The revocation process can be triggered by also the \emph{Domain Owner}s.  Moreover, the revocation status of all the TLS certificates is transparent, can be tracked and verified in \emph{CertLedger}. \newline

\textbf{Property 3:} \emph{CertLedger eliminates certificate revocation checking problems due to non-existent, improper or malicious CRL/OCSP services offered by the CAs.} 

\emph{CertLedger Client}s do not need \emph{CA}s to check the revocation status of the TLS certificates. They receive the final state of the TLS certificate with its proofs from the domain during the TLS handshake. They can verify the revocation status of the TLS certificate using the state tree hash in the block headers. Thus, CRL and OCSP services of the \emph{CA}s are not needed to make revocation check.  \newline

\textbf{Property 4:} \emph{CertLedger preserves the privacy of the clients during TLS handshake.}

While connecting to a domain during the TLS handshake, the \emph{CertLedger Client}s can check the authenticity and the revocation status of a TLS certificate without requiring any further network connection, hence their privacy is fully preserved. \newline

\textbf{Property 5:} \emph{Trusted CA Management in CertLedger is transparent and can be publicly verified.}

In \emph{CertLedger},  the requirements of being a Trusted CA is defined in the open standards. Only the CAs complying these standards are accepted as a Trusted CA upon inspection of the \emph{CertLedger Board Member}s. The number of required Board Members for a valid inspection is specified in a threshold manner. The inspections are conducted periodically. 

Fraudulence and malfunctioning reports about Trusted CAs are also managed in the \emph{CertLedger}. Any \emph{CertLedger Client} who has detected a fake but valid TLS certificate, can put all the fraudulency proofs in the \emph{CertLedger}. The accused Trusted CA can also put his justifications in the \emph{CertLedger}. Upon inspection of the proofs, \emph{CertLedger Board Member}s can change the status of the Trusted CA as 'Untrusted' in a threshold manner.\newline

\textbf{Property 6:} \emph{CertLedger eliminates Certificate/Key Stores in TLS clients.}

Management of Trusted CA root certificates and public keys of logs is a burden and one of the main source of security compromise\cite{CAMan1,CAMan2}. At the moment, TLS clients  store them in their Certificate/Key Stores which can be more than one with different contents for a client. However, \emph{CertLedger Client}s do not need to manage the Trusted CA certificates and public keys  in a variety of Certificate/Key Stores. Namely, the trustworthiness and the integrity of these certificates and keys are assured by the blockchain architecture of the \emph{CertLedger}. \newline

\textbf{Property 7:} \emph{CertLedger eliminates requirement for external Log Auditors.}

\emph{CertLedger} do not require external auditing for checking its cryptographic consistency and behaviour since new blocks are verified and appended to the log only upon the consensus of its clients. \newline

\textbf{Property 8:} \emph{CertLedger provides only one single implementation for certificate validation process.}

Improper implementation of TLS certificate validation process by software vendors is one of the top ten OWASP security vulnerabilities \cite{ImproperCertVal} in 2017. It is a common issue with one internet crawler to be able to connect to a TLS protected domain while encountering connection problems with another crawler due to differences in certificate validation processes. \emph{CertLedger}  eliminates this issue by validating TLS certificates prior to appending to the ledger according to the international standards stated in Section \ref{architecture}. Therefore, \emph{CertLedger Client}s do not need to validate a TLS certificate during a TLS handshake, they only check its existence and revocation status in the \emph{CertLedger}. \newline

\textbf{Property 9:} \emph{CertLedger minimizes attack duration due to instant monitoring.}

\emph{CertLedger} does not prevent issuance of fake but valid TLS certificates and require monitoring. However, \emph{CertLedger} provides infrastructure in order to detect fake TLS certificates and fraudulent revocations instantly.   Mechanisms such as event watchers can be implemented which listen \emph{CertLedger} events and inform a \emph{Domain Owner} through several means (e.g., e-mail, SMS) upon a change  on the state of his TLS certificates. The events in smart contract based blockchain architectures can easily be watched through APIs such as web3 \cite{web3} for Ethereum. Therefore, in \emph{CertLedger}, monitoring process is effortless and the attack duration can be minimized due to promptness. \newline

\section{Performance of CertLedger} \label{performance}

\subsection{Storage}

According to Verisign domain name industry brief, there are about $3.3 \times  10^8$ registered domain names at the end of 2017 \cite{DomainNum} and approximately half of them are using TLS certificates \cite{SSLDomainNum}. For the calculations we consider the TLS certificates with elliptic curve (EC) cryptography, since they will be dominantly used across the internet due to smaller key sizes, higher performance, and increased security. The size of a TLS certificate which uses 256 bits of EC public key, signed with ECDSA signature algorithm, is approximately $2^9 $ bytes. 

The total disk space required to store all of the TLS certificates is roughly about $ 1.65 \times 10^8 \times 2^9  bytes \approx 84 $ GB. The maximum life time period of TLS certificates is specified  as two years (825 days) in a recent CAB Forum Ballot \cite{CertLifeTime}. 

Note that the size of the underlying blockchain of the \emph{CertLedger} depends on its block size and the frequency of adding new blocks. Block size increases with respect to the number of comprised transactions and the target block time depends on the selected consensus algorithm of the \emph{CertLedger}. (i.e., the target block time in Bitcoin PoW consensus algorithm  is a tradeoff between the propagation time of the new blocks and the amount of work wasted due to chain splits,  and the target block time in the Ouroboros algorithm  is a simple parameter that can be changed at any time according to the network efficiency\cite{Bitcoinwiki,kiayias2017ouroboros}). While making the calculations, for illustration purposes target block time is selected as 10 minutes considering the following aspects. 

\begin{itemize}
	\item The required storage capacity of the \emph{CertLedger Client}s will increase with shorter block times where some of them have storage constraints and cannot be upgraded easily.
	\item Selected block time should be discouraging for an adversary to make an attack.
\end{itemize}

Moreover, we also estimate that the transactions will be about mostly adding new certificates where unnecessary revocations will be avoided due to the cost of new TLS certificates. In this respect, storage space requirements of a \emph{Full Node} for keeping the transactions of the \emph{CertLedger} and \emph{CertLedger Client} can be detailed as follows: 

\begin{itemize}
	\item \textbf{Full Nodes:}

	


	If we assume the average TLS certificate lifetime will be one year, \emph{Full Nodes}  need approximately 84 GB of disk space per year  for storing the TLS certificates (i.e., total storage of miners= 84 GB $\times$ year).  Considering the fact that the price of a 1 GB disk storage is approximately 0.02 \$ \cite{diskprices}, the cost of storing \emph{CertLedger} transactions for a non-mining full node is about 3.4 \$, which is not a deterrent amount. 
	
	
	These calculations are based on the assumption of storing the certificates for all the TLS protected domains. However, the required disk space can be decreased during the initial phase of the \emph{CertLedger}  by only adding the certificates of the domains which have high risk for user privacy and confidentiality (e.g., finance, social media, shopping, mailing) and adding the rest in course of time.

	\item \textbf{CertLedger Clients:} 
	They do not store the blocks but only the headers, and their required disk space is directly proportional with the stored number of blocks. As Satoshi describes in \cite{nakamoto2008bitcoin}, the required disk space to store block headers in Bitcoin for 10 minutes block interval time is 4.2 MB ($6 \times 24 \times 365 \times 80$ byte)  per year. A similar calculation for Ethereum with the same block internal time results in 26 MB of disk space per year. We expect \emph{CertLedger} header size to be similar to Ethereum. In this respect, for a two years of period, light node \emph{CertLedger Client}s will require approximately 50 MB of disk space for all TLS domains \footnote{There are also other studies proposed for efficient light node clients in recent studies such as NIPoPoWs by IOHK \cite{kiayias2017non} and FlyClient by Luu et al.  \cite{FlyClient}. For instance, FlyClient proposes, not only storing previous blocks hash in block headers, but also the root of a merkle tree which commits to all blocks. Therefore, a logarithmic sized proof will be enough to verify whether a specific block is a part of the blockchain. Consequently, it is sufficient for the light nodes to store the head of the chain. They only require another Merkle proof to verify whether a transaction is included in the block.}.

\end{itemize}

\subsection{Communication}

\emph{CertLedger} full node clients do not need to make network connection to verify a TLS certificate during TLS handshake. However, due to underlying consensus architecture, they need to exchange all the generated transactions and the blocks with their peers in the P2P network.

\emph{CertLedger} light node clients need to receive the final state of the \emph{Domain State Object} and the related Merkle proofs for the state tree in TLS extensions from the domain during the TLS connection. These extensions will bring an overhead to the communication between the client and the domain. As a part of the P2P network, they only download block headers from the P2P network, but do not exchange blocks and transactions with other peers as the light nodes in other blockchain architectures such as Bitcoin and Ethereum.

\subsection{Computation}

Lookup, insertion, and update operations depend on the data structures used in the \emph{CertLedger}. Since \emph{CertLedger Client}s requires to verify the final state of the TLS certificates efficiently, \emph{CertLedger} can use modified Patricia trees as in Ethereum in order to process the aforementioned operations in $O(log(n))$ where $n$ denotes the number of TLS certificates \cite{PatriciaTrie}.

\section{Comparison} \label{comparison} 

We now compare the \emph{CertLedger} with the previous proposals according to following criterias which is tabulated in Table \ref{tab:comparison}.  

\textbf{Log Proofs:}
\emph{CertLedger} provides proof for the existence and revocation status for all the TLS certificates as \emph{DTKI}. \emph{CT} provides proof only for the existence of a certificate in the log whereas  \emph{AKI}  and \emph{ARPKI} provide proofs only for the valid TLS certificates which are not revoked and expired. \cite{WLCWJZ18} provides existence proof for all of the TLS certificates, but may not provide a revocation proof if a CA does not issue correct CRL or gives OCSP service.

\textbf{TLS Handshake Performance:}
In  \emph{CT},  the proofs are sent to the clients during TLS handshake. However,  the received proof is not enough to make a  successful TLS handshake since TLS clients have to check the revocation status of the TLS certificate  by conventional methods like CRL or OCSP which can be cumbersome due to big CRL files and latency or interruption in OCSP services. Moreover,  if the TLS clients make a prior audit to the logs to check the existence of the proofs, handshake duration increases.  In \emph{DTKI}, TLS clients have to make network connections to both \emph{MLM} and the \emph{CLM} to request the proofs. In \cite{WLCWJZ18}, web servers send the certificate transactions and the Merkle proofs in TLS extensions to the browsers, and the browsers validate the transactions using the block headers. In addition, browsers with high security concerns may also spend time for revocation checking with conventional methods.

Full node \emph{CertLedger Client}s  do not require any external proofs for a successful TLS handshake and can check the existence and the revocation status of a TLS certificate locally, but light node clients have to receive and verify the final state of the TLS certificate and the related \emph{Merkle Audit Proof} from the domain.

\textbf{Independent Logs:} In \emph{CT}, \emph{AKI} and \emph{ARPKI}, there can be independent logs which comprise different sets of TLS certificates. If a TLS client is trusting to a log and the visited domain's TLS certificate has not been appended to the log, then the TLS connection will be unsuccessful.  For the CAs and the domain owners, it is impossible to know  the TLS clients' set of trusted logs.  Hence, ideally, CAs have to append their TLS certificates to all of the independent logs to eliminate the unsuccessful TLS connections. But the set of logs is not fixed and new logs can arise anytime. Appending TLS certificates to a changing set of logs is a burden for the CAs.  Moreover, necessity of monitoring and auditing  different logs is another overhead which can also lead to security compromise.

However, in \emph{CertLedger}, all of the TLS certificates are appended to one single log with multiple copies, and \emph{CertLedger Client}s do not need a trusted key for the verification of the log proofs.  Only this single log have to be monitored.

\begin{table*}
	\caption{Comparison of Log Based Approaches to Certificate Management}
	
	\label{tab:comparison}
	\footnotesize
	\scalebox{0.72}{
		\begin{minipage}{28cm}
			
			\begin{tabular}{|>{\bfseries}l|c|c|c|c|c|c|}
				\cline{2-7} 
				\multicolumn{1}{c|}{} & \textbf{CT} \cite{RFC6962} & \textbf{AKI\cite{Hyn2013}} & \textbf{ARPKI\cite{Basin:2014}} & \textbf{DTKI\cite{YuCR14}} & \cite{WLCWJZ18}&\textbf{CertLedger} (This paper)\tabularnewline
				\hline
				\makecell[l]{External Dependency During\\TLS Handshake} & Yes & Yes & Yes\footnote{Clients have to make standard certificate validation which may require external resources}  & Yes & No& No \tabularnewline
				\hline 
				\makecell[l]{Factors Effecting \\ TLS Handshake Performance}  & \makecell[l]{Certificate and proof \\verification, \\Revocation  checking,\\Auditing (optional)}  & \makecell[l]{Certificate and proof \\verification,\\Occasional  checks\\ for the ILS root \\ hash with the\\ \emph{Validator}s} &  \makecell[l]{Certificate and proof\\ verification}&   \makecell[l]{Certificate and proof\\ verification,\\Connection to\\ MLM and CLM}  & \makecell[l]{Blockchain update,\\Verification of the\\ transaction and\\ the proof }  & \makecell[l]{Data transmission\\ overhead and\\verification of the\\ Merkle audit proof} \tabularnewline
				\hline
				\makecell[l]{Existence of Logs with\\Different Contents}  & Yes & Yes & Yes  & No & No &No \tabularnewline
				\hline
				\makecell[l]{Necessity of External \\ Auditing} & Yes & Yes  & Yes & Yes  & No &No \tabularnewline
				\hline
				\makecell[l]{Necessity of External \\ Monitoring} & Yes & Yes  & Yes &  No \footnote{Makes an unrealistic assumption that fake Master certificates cannot be issued, apparently not relevant for strong adversaries}& Yes & Yes\footnote{Monitoring is efficient, it is possible to be aware of the TLS certificate state change promptly}\tabularnewline
				\hline
				\makecell[l]{Does TLS Clients Require\\to Store Trusted Keys\\or Certificates of Logs ?}  & Yes & Yes & Yes & Yes & Yes& No  \tabularnewline
				\hline

				\makecell[l]{TLS Clients Requires to Store\\ a copy of the Log} & No & No & No & No & Partly (block headers) &\makecell[c]{ Partly (block headers)}   \tabularnewline			
				\hline
			\end{tabular}

		\end{minipage}
	}
	
\end{table*}

\textbf{Log Availability:} 
In \emph{CT}, logs have to be audited and monitored continually for consistency and finding fake TLS certificates.

In \emph{A(RP)KI}, after each \emph{ILS} update or  a time out period, domain owners need to download proofs from the \emph{ILS}. 
In \emph{DTKI}, clients have to request proofs from the \emph{MLM}, \emph{CLM} or their mirrors during each TLS connection. Due to these reasons, all of these proposals require the logs to be permanently available which may lead to a single point of failure. 
However, \emph{CertLedger} is maintained in a P2P network upon consensus of its clients. Every full node \emph{CertLedger Client} have a copy of the log and do not have a dependency to verify the TLS certificates. The light node clients receive the \emph{Domain State Object} of the TLS certificate and its merkle proofs from the domain and verify it locally using block headers. 

\textbf{Auditing and Monitoring:}
\emph{CT} requires to be audited by external auditors such as TLS clients and monitored by the CAs and the domain owners. In \emph{AKI}, there are \emph{Validator}s which check the consistency of \emph{ILS}s and whether the TLS certificates are updated according to certificate policies. In \emph{ARPKI}, \emph{Validator}s are optional and their role is distributed to other CAs or \emph{ILS}s. \emph{DTKI} relies on TLS clients for auditing, but do not require monitoring due to existence of master certificates. Expiration of master certificates and issuance of fake master certificates by the compromised CAs are open issues in \emph{DTKI} both related to monitoring. \cite{WLCWJZ18} does not require external auditing, however it does not propose an efficient monitoring architecture.

As already discussed in Section \ref{securityanalysis},  \emph{CertLedger} does not require external auditing, and can be monitored easily and promptly. 

\textbf{Key/Certificate Store Management:}
In \emph{CT}, TLS clients have to store Trusted CA certificates and the public keys of the logs to validate the TLS certificates and the \emph{SCT}s. Although, \emph{AKI} states that  installing the trusted CAs and \emph{ILS}s certificates and public keys to TLS clients will be enough to make a successful TLS connection, TLS clients also receive signed validator information from the domains during the TLS handshake. In order to validate this information, clients must also trust the \emph{Validator}s' public keys. In \emph{ARPKI}, TLS clients have to store the trusted CAs and \emph{ILS}s certificates to validate the proofs received from the domain. In \emph{DTKI}, TLS clients require only the \emph{MLM} public key to verify the queried proofs during a TLS connection. \cite{WLCWJZ18} clients continue to make the conventional trusted key/certificate store management.

However, Trusted CA certificates are stored and managed in \emph{CertLedger}.  While adding TLS certificates  to the \emph{CertLedger}, they are verified whether they are issued by one of the Trusted CAs. Since \emph{CertLedger Client}s do not  make further verification, but only check the revocation status of the TLS certificate during TLS handshake, they do not require to store and manage any trusted key or certificate.


\textbf{Privacy:} In \emph{CT}, TLS clients audit the log to ensure its consistency by requesting \emph{Merkle Audit Proof}s for the \emph{SCT} they have received from the domains during the TLS connection. Moreover,  they can use OCSP to check the revocation status of a TLS certificate. These processes enables logs and the OCSP servers, to trace the browsing history of the TLS clients. 
In \emph{AKI} and \emph{ARPKI}, domains send the proofs to the TLS clients, thus their privacy is preserved.
In \emph{DTKI}, TLS clients both query \emph{MLM} and the \emph{CLM} to receive proofs for the targeted domain during TLS handshake. It is proposed that, to preserve privacy, the domains can act as a proxy to make these queries. However, this proposal will increase the network latency even more in DTKI. TLS clients in \cite{WLCWJZ18} continue to make conventional revocation checking, hence their browsing history can be traced  by the OCSP servers.

In \emph{CertLedger}, as discussed in Section \ref{securityanalysis},  privacy of the clients are fully preserved. 

\section{Conclusion and Future Work} \label{conclusion}

There have been serious security incidents due to corrupted CAs which issued fake but valid TLS certificates. To make CAs more transparent and to verify their operations, public logs are proposed in recent studies such as CT, AKI, ARPKI and DTKI.  CT is proposed by Google which has almost half of the browser market share. Google  made CT mandatory in Chrome for all issued TLS certificates after April 2018 \cite{TLSHistory}. However, CT and the other proposals are subject to split-world attacks.

We propose a new PKI model with certificate transparency based on blockchain, what we called \emph{CertLedger}. In the \emph{CertLedger}  all TLS clients can verify the final state of log, which makes split-world attack impossible. Moreover, \emph{CertLedger} also provides transparency in certificate revocation and trusted CA management processes.

A future work would be implementing \emph{CertLedger} in an existing blockchain framework (e.g., Ethereum or Neo) and make a detailed performance and usability analysis.  Moreover, in order to eliminate the \emph{CertLedger Board Member}s, introducing an automated mechanism running without human intervention will increase the transparency of the trusted CA management.

\nocite{*}

\bibliographystyle{compj}
\bibliography{CertLedger}

\end{document}